# The interplay between exciton- and phonon-induced superconductivity might explain the phenomena observed in LK-99

Junhui Cao[1,2] and Alexey Kavokin[1,2,3]

[1]Westlake University, School of Science, 18 Shilongshan Road, Hangzhou 310024, Zhejiang Province, China

[2]Westlake Institute for Advanced Study, Institute of Natural Sciences, 18 Shilongshan Road, Hangzhou 310024, Zhejiang Province, China

[3]Spin-Optics laboratory, St. Petersburg State University, St. Petersburg 198504, Russia

Keywords: Superconductor, exciton, BCS theory, Meissner effect

The experimental results hinting at the room temperature and ambient pressure superconductivity and magnetic levitation in LK-99 attracted an unprecedented interest. While attempts of other teams to reproduce the reported observations on similar samples failed so far, it seems worthwhile to try building a theoretical model that would explain the ensemble of the available data. One of important features that needs to be explained is an apparent contradiction between an extremely high critical temperature $T_c$ and rather modest critical magnetic field $B_c$ and critical current $j_c$ reported for LK-99. We show theoretically, that these data may be quantitatively reproduced assuming the interplay between exciton- and phonon-induced superconductivity, while the conventional BCS or Brinkman-Rice-Bardeen-Cooper-Schriefer (BR-BCS) mechanisms would result in a much higher $B_c$ for the same $T_c$.

## 1. Introduction

The initial excitement induced by the reports on the room temperature and ambient pressure superconductivity [1] and magnetic levitation [2] by Lee *et al* was quickly replaced by disappointment caused by the failure of the first attempts of several other groups to reproduce the reported findings in samples of LK-99 prepared following the technology described by Lee *et al* and showing nearly identical X-ray patterns with their sample [3-7]. On the other hand, one could argue that it is very likely that polycrystalline samples produced in different labs may differ in their meso- and macroscopic structure, especially in what concerns the geometry of interface layers that seem to play a crucial role in the observed effects. There is no reason to assume that the reported experimental findings [1,2] are fake. More likely, the non-reproducibility of the data by other groups may be explained by a poorly studied and described fabrication technology. It might have happen than out of thousands of samples produced following by available guidelines only one or two would possess the morphology required to exhibit the reported electrical and magnetic properties. Still, the reported results are worth discussing. The claim of superconductivity has been supported by several sets of data showing the current-voltage dependencies taken at different temperatures and magnetic fields. These data allow to extract would be critical current and magnetic field as functions of temperature. As our analysis below shows these dependencies are rather non-trivial. Their possible interpretation with use of a gap equation is in the focus of this study. We publish it because the corresponding (may be strongly different) sets of experimental data will sooner or later be produced by other groups which are now involved in studies of LK-99. In this context, our theoretical analysis may appear to be timely and useful.

In what follows we compare the experimental temperature dependencies of the critical current and critical magnetic field with predictions of various models and conclude that neither Brinkman-Rice-Bardeen-Cooper-Schriefer (BR-BCS) nor flat band BCS superconductivity model, evoked by some recent theories [8], can qualitatively explain the observed dependencies. Namely, the experimental ratio of the critical magnetic field to critical temperature appears to be significantly lower than what one would anticipate. In order to resolve this controversy, we develop a model based on the interplay between the exciton- and the phonon-induced superconductivity [9-12]. We start from the concept of a superconducting quantum well (SQW) proposed by the authors of Refs. 1,2 and argue that an SQW may support stable dipole-polarized excitons formed by electrons from a quantization subband situated above the Fermi level and holes from the subband below the Fermi level. Scattering of conductivity electrons with dipole polarized excitons is expected to be more efficient than the electron-phonon scattering, and it may lead to Cooper pairing [11]. The interplay between the exciton-mediated and photon-mediated superconductivity may result in a strong increase of $T_c$ [12] while the critical magnetic field is much less affected and it remains relatively low [13]. We adapt the model of [12] to the specific experimental conditions of LK-99 and perform the numerical calculations that yield a reasonably good agreement with the experiment [1,2] with a minimum of free parameters. In particular, we reproduce in detail the experimental temperature and magnetic field dependencies of the critical current. To our knowledge, none of the previous analysis no matter supporting or negating superconductivity addressed these important experimental features that may be considered as a signature of LK-99. A good agreement between the predictions of the excitonic model developed here and the experiment might shed light on the origin of the observed phenomena.

## 2. Theory

In order to directly compare predictions of several models, we shall analyze the gap equation represented in the integral form [11]:

$$\Delta(\xi, T) = \int_0^\Omega \frac{U_0(\xi - \xi')\Delta(\xi', T) \tanh\left(\frac{E}{2k_B T}\right)}{2E} d\xi', \qquad (1)$$

where $\Delta$ is the superconducting gap, $k_B$ is the Boltzmann constant, $U_0$ is the effective electron-electron interaction potential, $\xi$ is the electron energy difference with respect to the Fermi energy, $E = \sqrt{\xi'^2 + \Delta^2(\xi', T)}$. The cutoff for the integral $\Omega$ is usually associated with the Debye frequency, that is the maximum frequency for the lattice vibrations. In the simplest approximation, $U_0(\xi)$ is assumed to be constant below the Debye frequency. The critical magnetic field $B_c$ and critical current density $j_c$ are linked with the temperature-dependent value of the gap function at zero energy by the relations [12]:

$$B_c = \sqrt{\mu_0 N(0)} \Delta(0), \qquad (2)$$

$$j_c = \frac{e N(0) \Delta(0)}{\hbar k_F}, \qquad (3)$$

where $\mu_0$ is the vacuum permeability, $N(0)$ the electron density of states near the Fermi surface, and $\hbar k_F$ the electron Fermi momentum.

In what follows we show that the experimental findings of Refs. [1,2] cannot be reproduced within the BCS theory [14] even if the unrealistically high electron-phonon coupling strength is assumed. In contrast, the interplay of exciton- and phonon-induced Cooper pairing might explain the observed phenomena. The exciton mechanism for electron-electron coupling was initially proposed for virtual excitons in hybrid metal-semiconductor heterostructures [9,10]. It has never been evidenced experimentally, to our knowledge, most likely because the coupling strength of spatially separated electrons and excitons is usually too weak. Here we shall take advantage of the updated version of this mechanism proposed by Laussy et. al. [11]. We shall assume that real excitons do exist in the structure and their wavefunctions strongly overlap with free electron (free hole) wavefunctions. This is likely to be the case in the context of a superconducting quantum well (SQW) that is formed due to local strains in LK-99 according to Ref. 1. Such a quantum well would be characterized by a set of electronic subbands, some of which would be fully occupied and some would be fully empty at zero temperature. The subband that contains the Fermi level would be partially occupied and it would mostly contribute to the conductivity.

At non-zero temperatures, a significant number of electrons would be situated in the first subband above the Fermi level, and a significant number of holes would be situated at the subband below the Fermi level. If the binding energy of excitons formed by such electron-hole pairs exceeds the energy splitting between relevant subbands, these excitons would be stable even at zero temperature, and a sort of the excitonic insulator (Mott insulator) phase would be formed in the plane of SQW. This excitonic insulator could coexist with a free electron (hole) gas in a partly filled subband. Scattering of excitons with electrons from this subband would trigger an excitonic mechanism of Cooper pairing qualitatively similar to the familiar phonon-induced Cooper pairing (see the diagram in Fig. 1).

The differences of the exciton-induced and phonon-induced Cooper pairing are in the energy dependences of the corresponding coupling potentials. Compared with the electron-phonon interaction, the electron-exciton interaction might be much stronger as it is governed by the Coulomb interaction of electrons with dipole-polarised excitons in the same SQW. The energy dependence of the exciton contribution to the electron-electron interaction potential would be governed by the density of excited exciton states that is expected to be staircase-like due to the multitude of subbands where excitons may be scattered to.

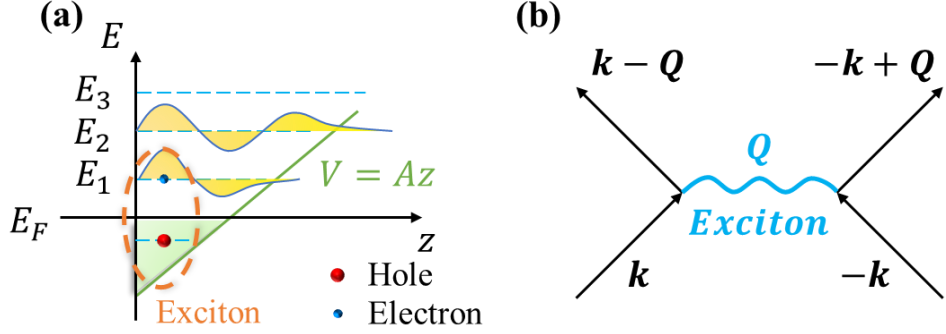

Figure 1. (a) Confinement of excitons in a triangular superconducting quantum well (schematic). The horizontal dashed lines indicate the quantization of energy levels, with the wavefunction (Airy function) shown above. The first three levels of electrons above Fermi energy are denoted as $E_{1,2,3}$. (b) A diagram of the exciton mediated electron-electron scattering, showing the momentum $\boldsymbol{Q}$ transfer by the exciton mechanism.

In order to evaluate the exciton-induced electron-electron interaction potential let us consider a model SQW that might be formed due to the local strain in LK-99. It is well known that the electronic band bending occurs at the interface of two different materials with a disparity of chemical potential, resulting in the formation of a built-in electric field at the surface or interface [15]. In most part of cases, such as the AlGaAs/GaAs heterojunctions and organic interfaces [16], the bending effect leads to the formation of a nearly triangular potential well (Fig.1a) [17, 18], which induces the energy quantization of a two-dimensional electron gas (2DEG), whose Hamiltonian writes $H(z) = T + V(z) = -\frac{\hbar^2 \nabla^2}{2m} + Fz$. Ref. 1 argues that the similar potential is formed in LK-99 due to the replacement of some of led ions by copper ions. Figure 3e of Ref. 1 schematically shows this potential, while its exact shape remains unknown. Here, to be specific, we assume the steepness of the triangular potential $F = 0.5$ meV · nm$^{-1}$. This value is a parameter of the model. The wavefunctions of electrons in triangular quantum wells are Airy functions [19], given by

$$\varphi_i = \frac{\pi}{\sqrt{3}} \sqrt{\xi} \, [J_{\frac{1}{3}}(\frac{2}{3}\xi_i^{\frac{3}{2}}) + J_{-\frac{1}{3}}(\frac{2}{3}\xi_i^{\frac{3}{2}})], \qquad (4)$$

$$\xi_i = \left(\frac{2m_e}{\hbar^2 F^2}\right)^{\frac{1}{3}} (Fz - E_i),$$

where $m_e$ is the electron effective mass, $J_i$ are the Bessel functions, and $i \in \{1,2,3,...\}$ indicates the index of the quantization subband. An infinitely high barrier is assumed for $z < 0$. We shall consider the excitonic scattering between three subbands whose energies are $E_1 = 16.87$ meV, $E_2 = 29.50$ meV, $E_3 = 39.84$ meV, respectively, see the scheme in Fig.2a. The scattering of a conductivity electron with an exciton kicks the corresponding exciton out from its ground sate ($k = 0$) of the first subband and brings it to an excited state ($k \neq 0$) of the same subband, or to one of the higher energy subbands. Note that the exciton scattering as a whole particle can be reduced to the simultaneous electron and hole scatterings. In the next estimations we shall evaluate the magnitude of the electron-exciton scattering assuming that it is dominated by electron-electron repulsion.

For the electron-exciton interaction in the first subband, Skopelitis et al [12] and Sedov et. al. [20] expressed the effective electron-bogolon coupling constant as $\eta_1 = \frac{X^4 L^2 d^2 e^4 n_p^2 A N(0)}{g_0 \varepsilon^2}$, where $X = 1$ is the Hopfield coefficient for exciton, $L = 20$ nm is the average electron-exciton spacing, $d = 10$ nm the exciton dipole length, $n_p$ the exciton density, $A$ is the normalization area, $N(0)$ is the free electron density of states at the Fermi surface, $g_0$ is the interaction constant, $\varepsilon$ is the dielectric permittivity of the background. We estimate $X = 1$, $L = 5$ nm, $d = 7$ nm, $n_p = 5 \times 10^{10}$ cm$^{-2}$, $N(0) = \frac{m_e}{\pi \hbar^2}$, $\frac{g_0}{A} \sim 3$ meV [21] and $\varepsilon = 6\varepsilon_0$. According to the Fermi golden rule, the probability of an electronic transition from the first subband to the second one is $P_{1 \to 2} = \int |\langle \varphi_2 | \varphi_1 \rangle|^2 D(E) \, dE$, where $\varphi_1$ and $\varphi_2$ are the wavefunctions given by Eq.4. For a two-dimensional electron gas (2DEG), the electron density of states $D(E) = \frac{m_e}{\pi \hbar^2}$ is independent of energy. The ratio of electron scattering probabilities from the first to the second subband and within the first subband $\eta_2 = \frac{P_{1 \to 2}}{P_{1 \to 1}} \eta_1$. Where $\frac{P_{1 \to 2}}{P_{1 \to 1}} = \frac{|\langle \varphi_2 | \varphi_1 \rangle|^2}{|\langle \varphi_1 | \varphi_1 \rangle|^2} = 0.7466$. The cut-off frequency of the electron-electron interaction channels corresponding to the exciton scattering within the first and the second subband must be close to the energy splittings between the corresponding subbands. We shall assume $\Delta_1 = E_2 - E_1 = 12.63$ meV, $\Delta_2 = E_3 - E_2 = 10.34$ meV, respectively, for the first two steps of the interaction potential. We shall neglect steps caused by the exciton scattering to the third, firth etc subbands as the probability of such scatterings would be very low.

In this approximation, the total effective electron-electron interaction potential profile is composed by the two-step electron-exciton interaction potential with magnitudes $\eta_{1,2}$ and the comparatively lower magnitude electron-phonon interaction potential having a cut-off at the Debye energy, as shown in Fig.2b.

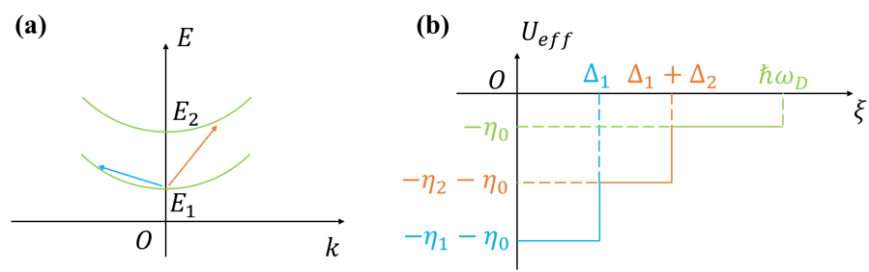

Figure 2. (a) The subband structure of the considered model SQW. Blue and orange arrows indicate intra/inter-subband scattering channels. (b) Three-step effective electron-electron interaction potential profile. $\eta_{0,1,2}$ are the magnitudes of the components of the interaction potential due to the electron-phonon, electron-exciton (intra-subband) and electron-exciton (inter-subband) scatterings, respectively.

## 3. Results and discussions

We shall be solving the gap equation (1) by the iteration method, accounting for the hybrid exciton-mediated and phonon-mediated electron-electron interaction potential shown in Fig. 2(b). We set the Debye temperature at 500 K that is close to the experimental values reported in Ref. 1. The effective electron-electron interaction constants are evaluated as $\eta_1 = 4.00$, $\eta_2 = 2.98$, $\eta_0 = 0.30$. Fig.3 shows the results of our model calculation compared to the experiment data of Lee *et al* [1]. We find a reasonably good agreement between the theory and the experiment. In Fig.3a, the critical current reaches about 250 mA at 300 K, then it monotonically decreases with the temperature increase, in full agreement with the experiment. The critical current dependence on the magnetic field given by Eqs. (1-3) is nearly linear and it repeats the experimental data very closely.

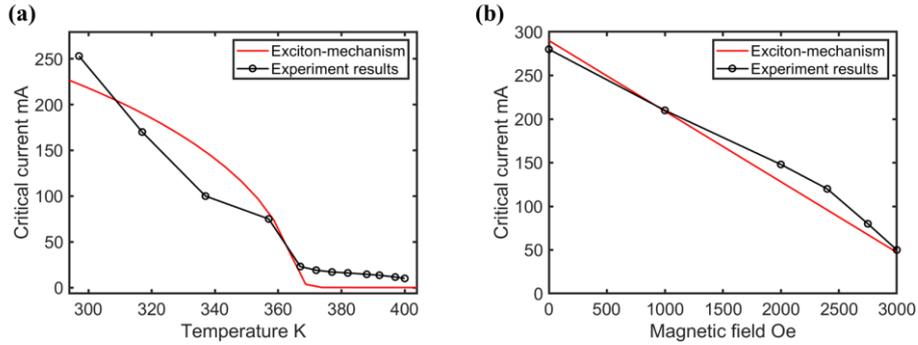

Figure 3. (a) Dependence of the critical current on temperature, and (b) shows the dependence of the critical current on magnetic field at T=270 K. The red solid curves show the results of our model calculation, while the points are experimental data adapted from [1] with the black lines being guides for an eye.

The results of our calculation are compared with predictions by the conventional phonon-BCS theory and by the BR-BCS model brought up by Hyun-Tak Kim [22] (see the Appendix A). In the calculations within the BCS and BR-BCS models, we have intentionally tuned up the electron-electron interaction strength in order to fit $T_c$. For the BCS model calculation, the electron-phonon coupling constant is set to $\eta_0 = 3.2$, which is unrealistically large, unless we deal with the flat-band superconductivity. Fig.4a shows the calculated temperature dependences of the superconducting gap $\Delta(0)$. One can see that the model developed in the present work yields the lowest gap as compared to the BCS and BR-BCS models. At zero temperature, our model yields the gap of 43.69 meV, while the BCS theory and the BR-BCS theory predict the gap energies of 60.79 meV and 55.37 meV, respectively. The inset shows the prediction of the phonon BCS model assuming a realistic electron-phonon coupling strength and electronic density of states (no flat bands). It yields the zero-temperature gap value of only 0.03 meV.

Figure 4(b) shows the critical magnetic field calculated using Eq. (2) as a function of temperature within the three above mentioned models. One can see that our present model yields the lowest value of the critical magnetic field in a wide temperature range: from zero till the room temperature. This is not surprising, as our model predicted the lowest superconducting gap. Figure 4(c) shows the critical current calculated as a function of the magnetic field at the room temperature. It is worth noticing that both the critical magnetic field and critical current obtained within our model are the smallest among all models. The critical magnetic field is 3000 Oe (0.3

Tesla) and the critical current is 300 mA at 290 K. These results are in agreement with the experiment, while the calculations within the BCS and BR-BCS models deviate from the experiment quite strongly.

We explain this difference between the predictions of three models by the interplay of two mechanisms of superconductivity resulting in a specific three-step interaction potential within our model. The non-trivial energy dependence of the interaction potential leads to the smoother dependence of $\Delta(0)$ on temperature as compared to the other models. Due to this smooth temperature dependence of $\Delta(0)$, very high critical temperatures may coexist with relatively modest critical magnetic field and critical current.

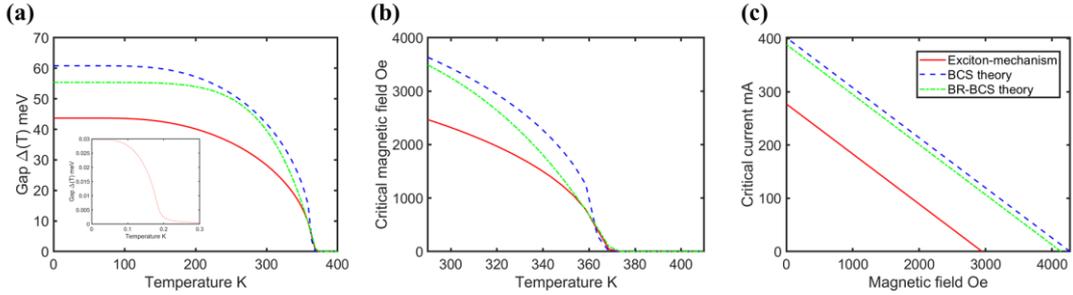

Figure 4. (a) The temperature dependence of the superconducting gap, (b) the temperature dependence of the critical current,(c) the critical current vs magnetic field at T=290 K. The red solid curve, the blue dashed curve and the green dash-dotted curve show the results obtained from within the model developed in this work, the conventional BCS theory and the BR-BCS theory, respectively. The inset to Figure 4(a) shows the temperature dependence of the superconducting gap obtained within the BCS model assuming a realistic value of the electron-phonon coupling constant.

## 4. Conclusion

In conclusion, we have reproduced the experimental dependencies of the critical current on temperature and magnetic field in LK-99 within the model assuming an interplay between exciton- and phonon-induced superconductivity in superconducting quantum wells. We have also checked that the conventional BCS model and the BR-BCS model evoked in Ref. 2 would lead to the critical current and critical magnetic field strongly exceeding the experimental values. Clearly, our estimations are inexact as we do not know that exact shape of the superconducting quantum wells in LK-99. It is no more than the hypothesis. Still, it allows one to explain the apparent controversy between the extremely high value of $T_c$ reported in Refs. [1,2] and relatively modest critical current and magnetic field values.

We would like to note at this point that, while the exciton-induced superconductivity is a theoretical concept that still lacks experimental confirmation, it is worth attention of the community in view of a rapidly expanding class of systems and materials where free electrons and excitons may coexist. Clearly, when the concept was brought forward by Bardeen and Ginzburg that was not the case. One should expect that the excitonic mechanism will strongly manifest itself sooner or later, which justifies our effort to analyse its potential impact on the temperature dependencies of critical current and critical magnetic field.

Finally, we would like to emphasize that we do not consider one-dimensional superconductivity in this work. Rather, we believe that two-dimensional superconducting layers may be formed in the interface regions of polycrystalline LK-99 sample. Two-dimensional superconductivity could be responsible for the reported SQUID data [1]. A further experimental analysis of the morphology of LK-99 samples and spatial location of insulator and superconductor phases is needed in order to conclude on the validity of our analysis.

We thank Professor Anvar Zakhidov for many enlightening discussions.

**References:**

[1] Sukbae Lee, Ji-Hoon Kim, Young-Wan Kwon, The First Room-Temperature Ambient-Pressure Superconductor, arXiv:2307.12008 [cond-mat.supr-con] (2023).

[2] Sukbae Lee, Jihoon Kim, Hyun-Tak Kim, Sungyeon Im, SooMin An, and Keun Ho Auh, Superconductor Pb10-xCux(PO4)6O showing levitation at room temperature and atmospheric pressure and mechanism, arXiv:2307.12037 [cond-mat.supr-con] (2023)

[3] Qiang Hou, Wei Wei, Xin Zhou, Yue Sun, Zhixiang Shi, Observation of zero resistance above 100 K in Pb10-xCux(PO4)6O, arXiv: 2308.01192v1.

[4] Lowe, Derek (1 August 2023). "A Room-Temperature Superconductor? New Developments". Chemical News. *"In the pipeline"* (blog). American Association for the Advancement of Science. Archived from the original on 1 August 2023. Retrieved 1 August 2023 – via Science.org

[5] Li Liu, Ziang Meng, Xiaoning Wang, Hongyu Chen, Zhiyuan Duan, Xiaorong Zhou, Han Yan, Peixin Qin, Zhiqi Liu, Semiconducting transport in Pb10-xCux(PO4)6O sintered from Pb2SO5 and Cu3P, arXiv:2307.16802 (2023).

[6] S. Zhu, W. Wu, Zh. Li, J. Luo, First order transition in Pb10-xCux(PO4)6O (0.9<x<1.1) containing Cu2S, arXiv:2308.04353 (2023).

[7] Dan Garristo, LK-99 isn't a superconductor — how science sleuths solved the mystery, Nature *doi: https://doi.org/10.1038/d41586-023-02585-7.*

[8] S.M. Griffin, Origin of correlated isolated flat bands in copper-substituted lead phosphate apatite, https://arxiv.org/abs/2307.16892, (2023); Junwen Lai, Jiangxu Li, Peitao Liu, Yan Sun, and Xing-Qiu Chen, First-principles study on the electronic structure of Pb10−xCux(PO4)6O (x=0, 1), arXiv: 2307.16040v1 (2023).

[9] D. Allender, J. Bray, and J. Bardeen, Model for an Exciton Mechanism of Superconductivity, Phys. Rev. B 7, 1020 (1973).

[10] V. L. Ginzburg, High-temperature superconductivity–dream or reality? Usp. Fiz. Nauk. 118, 315 (1976) [Sov. Phys. Usp., 19, 174 (1976)].

[11] F. P. Laussy, A. V. Kavokin, and I. A. Shelykh, Exciton-polariton mediated superconductivity, Phys. Rev. Lett. 104, 106402 (2010).

[12] P. Skopelitis, E.D. Cherotchenko, A.V. Kavokin and A. Posazhennikova, Interplay of Phonon and Exciton-Mediated Superconductivity in Hybrid Semiconductor-Superconductor Structures, Phys. Rev. Lett. 120, 107001 (2018).

[13] A. Kavokin and A. Zakhidov, Comment on: The interpretation of room-temperature superconductivity experiments in LK-99, Materials Today. *doi: https://doi.org/10.1016/j.mattod.2023.08.008*.

[14] J. Bardeen, L. N. Cooper, and J. R. Schrieffer, Theory of Superconductivity, Phys. Rev. 108, 1175 (1957).

[15] R. Dingle, H. L. StOrmer, A. C. Gossard, and W. Wiegmann, Electron Mobilities in Modulation-Doped Semiconductor Heterojunction Superlattices, Appl. Phys. Lett. 33, 665-667 (1978).

[16] P. Zhang, H. Wang and D. Yan, Dramatically Improved Electron Transport Performance by A Deep Triangular Potential Well in Organic Field-Effect Transistors, J. Phys. D: Appl. Phys. 53, 01LT01, (2020).

[17] A. N. Khondker and A. F. M. Anwar, Analytical Models for AlGaAs/GaAs Heterojunction Quantum Wells, Solid-State Electronics 30, 847-852 (1987).

[18] A. Mahajan1 and S. Ganguly, An Analytical Model for Electron Tunneling In Triangular Quantum Wells, Semicond. Sci. Technol. 36, 055012, (2021).

[19] J. Cao and A. Kavokin. Optical Properties of Magnetic Monopole Excitons. Condens. Matter, 8, 43 (2023).

[20] E. Sedov, I. Sedova, S. Arakelian, G. Eramo and A. Kavokin, Hybrid Optical Fiber for Light Induced Superconductivity, Scientific Reports, 10, 8131 (2020).

[21] F. Tassone and Y. Yamamoto, Exciton-Exciton Scattering Dynamics in A Semiconductor Microcavity and Stimulated Scattering into Polaritons, Phys. Rev. B 59, 10830 (1999).

[22] H. Kim, Room-Temperature-Super Conducting Tc Driven By Electron Correlation, Scientific Reports, 11, 10329 (2021).

**Appendix A**. BR-BCS theory

In the BR-BCS theory [22], the electron-electron correlations are taken into account as a correction to the traditional BCS theory. The critical temperature given by this model writes:

$$T_c = C(z)T_D^* e^{-\frac{\coth z}{\lambda_{BCS}^*}},$$

with $C(z) \equiv \frac{1}{2}\exp\left[-\coth(z)\int_0^x (\ln x\,/\cosh^2 x)dx\right]$, $z = \frac{T_D}{2T_c}$, $T_D^* = \rho^{\frac{1}{3}}T_D$ and $\lambda_{BCS}^* = A\lambda_{BCS}$ being the "renormalized" Debye temperature and electron-phonon coupling constant in the sense of Landau's Fermi liquid theory. The renormalization factor $A = \frac{N_0^*}{N_0} = \frac{\rho^{\frac{1}{3}}}{1-\kappa_{BR}\rho^4}$ is defined as the ratio of correlated electron density of states $N_0^*$ and free electron density of states $N_0$.

Recent calculations performed within the density functional theory have shown a flat band with the bandwidth of around 130 meV for LK-99 [8]. Based on this, we assume the renormalization factor $A$ to be around 40 for electrons in the 2D plane and $\kappa_{BR}$ to be 1. $\lambda_{BCS} = 0.035$, and $C(z) = 2.48$ for the Debye temperature of 500 K and the critical superconducting temperature of 370 K. These are the parameters assumed for the calculation within the BR-BCS model presented in this paper.